\documentclass[draftclsnofoot,11pt,journal,onecolumn]{IEEEtran}
\usepackage[top=1in,bottom=1in,left=1in,right=1in]{geometry}
\usepackage[utf8]{inputenc}
\usepackage[T1]{fontenc}
\usepackage{cite}
\usepackage{mathtools}
\usepackage{amssymb}
\usepackage{amsthm}
\usepackage{balance}
\usepackage{cprotect}
\usepackage{graphicx}
\usepackage{epstopdf}
\usepackage{algpseudocode}
\usepackage[usenames]{xcolor}
\usepackage[normalem]{ulem}
\usepackage{soul}
\usepackage{algorithmicx,algorithm}
\usepackage{fancybox}
\usepackage{tablefootnote}
\usepackage{mdframed}
\usepackage{wrapfig}
\usepackage{hyperref}
\usepackage{array}
\usepackage{tabularx}

\definecolor{ballblue}{rgb}{0.13, 0.67, 0.8}
\definecolor{blush}{rgb}{0.87, 0.36, 0.51}
\definecolor{chamoisee}{rgb}{0.63, 0.47, 0.35}
\definecolor{darkseagreen}{rgb}{0.56, 0.74, 0.56}
\definecolor{purple}{rgb}{0.5, 0.0, 0.5}

\newcommand{\revise}[1]{#1}

\mdfdefinestyle{MyFrame}{%
    linecolor=black,
    linewidth=2pt,
    leftmargin=0pt,
    rightmargin=0pt,
    backgroundcolor=gray!15,
    innertopmargin=4pt,
    innerbottommargin=4pt,
    innerrightmargin=4pt,
    innerleftmargin=4pt,
}

\title{A Guide to \revise{Computational Reproducibility} in\\Signal Processing and Machine Learning }
\author{Joseph~Shenouda and Waheed~U.~Bajwa}

\begin{document}
\maketitle

\bstctlcite{IEEEexample:BSTcontrol}

\section{Introduction}
\label{sec:intro}
\revise{A computational experiment is deemed \emph{reproducible} if the same data and methods are available to replicate quantitative results by any independent researcher, anywhere and at any time, granted they have the required computing power. Such \emph{computational reproducibility} is a growing challenge that has been extensively studied among computational researchers as well as within the signal processing and machine learning research community \cite{stoddenImplementingReproducibleResearch2014,raffStepQuantifyingIndependently2019a}. Signal processing research is in particular becoming increasingly reliant on computational experiments to test hypotheses and validate claims, which is in contrast to the yesteryears when one typically used computational experiments to elucidate rigorous theory and mathematical proofs. Therefore it has become more important than ever to ensure the reproducibility of computational experiments, as this is the first step in confirming the validity of research claims supported through the outcomes of computational experiments. But this is not turning out to be an easy task. The paradigm shift from the theory-driven research to the compute-driven claims in signal processing and machine learning has been facilitated by powerful computing resources, accessibility of massive datasets, and a myriad of new libraries and frameworks (such as NumPy \cite{harrisArrayProgrammingNumPy2020}, Scikit-learn \cite{pedregosaScikitlearnMachineLearning2011}, MATLAB Toolboxes~\cite{mathworksMATLAB}, and TensorFlow~\cite{abadiTensorFlowLargeScaleMachine}) that provide a layer of abstraction allowing for rapid implementation of complex algorithms. Unfortunately this changing research landscape is also bringing with it new obstacles and unseen challenges in developing reproducible experiments.}

Computational experiments today often incorporate various scripts for preprocessing data, running algorithms, and plotting results, all while utilizing huge datasets which require computing clusters that often take days or weeks to finish computing with multiple manual interventions needed to successfully produce the desired results. This is contrary to the way computational experiments used to be conducted and the way new researchers are introduced to computational resources in the classroom, where they typically use simple and intuitive interactive computing software consisting of a single script that runs locally on one's computer \cite{monajemiMakingMassiveComputational2016}. This new paradigm of computational experiments is now requiring the scientific community to rethink how we publicize and share our code to encapsulate all the necessary information about our experiments and \revise{make computational reproducibility practically possible}. Additionally our extensive dependence on libraries and frameworks leads to brittle codebases that typically only output correct results when executed on the original machine of the researcher. Furthermore, the nature of these data-driven experiments often require careful parameter tuning, random number generations, and data preprocessing, all of which are independent from the main finding, such as a new algorithm, being implemented or investigated.

\begin{mdframed}[style=MyFrame]
\begin{center}
    \includegraphics[width=15.5cm]{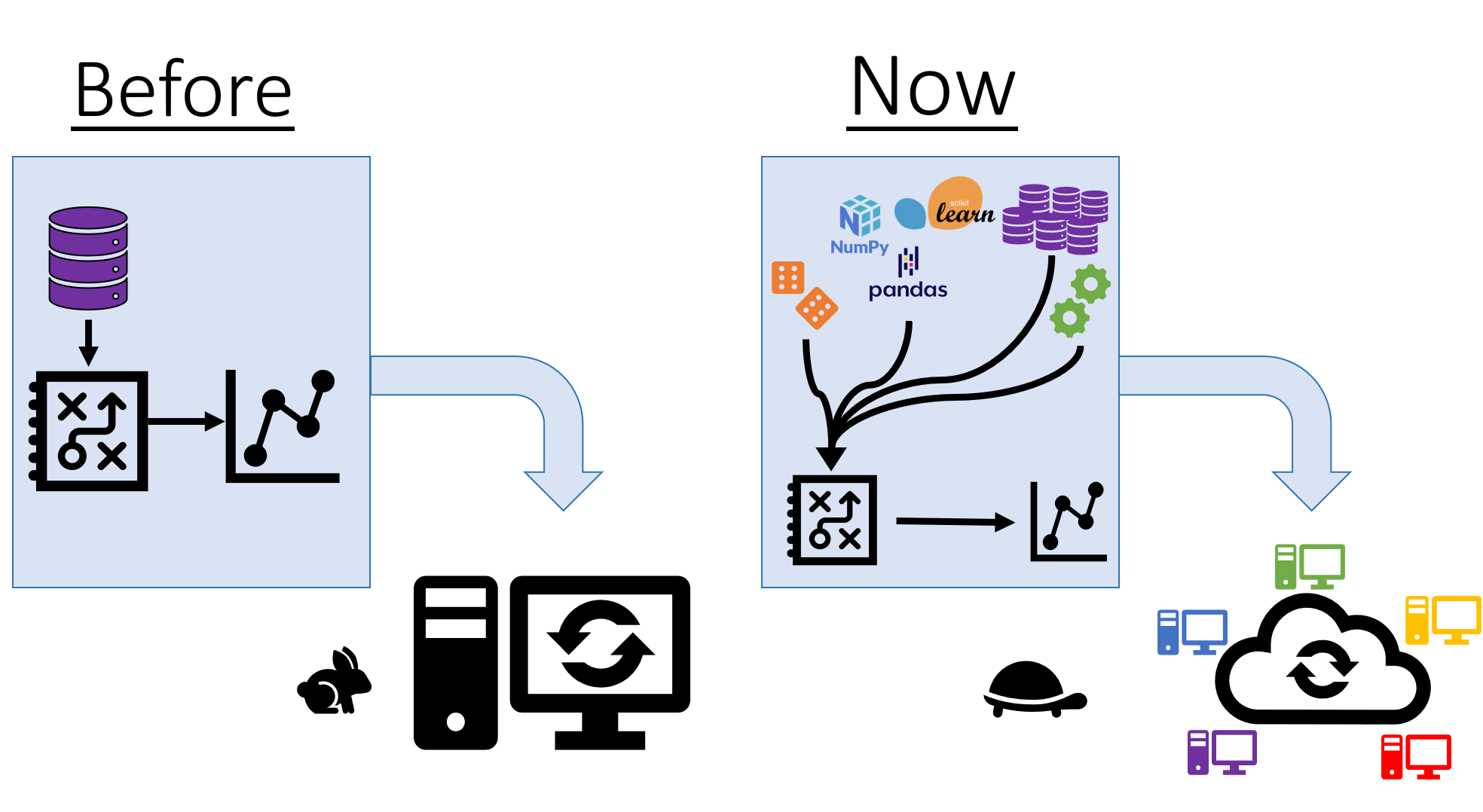}
\end{center}
\noindent\emph{Evolution of Computational Experiments}: Computational experiments in signal processing and machine learning today have transformed from the way they used to be conducted in the early days of these fields. Due to the rise in the availability of computational resources and large datasets, many of our \revise{computational} experiments can no longer be carried out on local workstations, as was the norm in the past. Rather, they are often carried out on large computing clusters and can take hours or days to complete. This fact, coupled with the dependencies on multiple third-party libraries, hyperparameter tuning, and random number generators makes it very time consuming, and sometimes nearly impossible, to try and reproduce published \revise{computational} results by trial and error.%
\end{mdframed}

\revise{Due to these new challenges most experiments have become difficult, if not impossible, to be reproduced by an independent researcher.} As an anecdote, when attempting to reproduce \revise{computational} results in our lab from a paper published just months prior, even the original authors of the experiment were unable to completely reproduce the results. However, this phenomenon is not unique to our lab. In 2016 a survey conducted by the journal Nature found that 50\%\ of researchers were unable to reproduce their own experiments \cite{baker500ScientistsLift2016}. And while the issue of reproducibility has been discussed in the literature \cite{stoddenImplementingReproducibleResearch2014} and specifically within the signal processing community \cite{vandewalleReproducibleResearchSignal2009,bjornsonReproducibleResearchBest2019}, it is still unclear to most researchers what are the \revise{most practical approaches} to ensure \revise{\emph{computational}} reproducibility without impinging on their primary responsibility of conducting research. \revise{This is because the guidelines and best practices provided for computational reproducibility in the existing works \cite{vandewalleReproducibleResearchSignal2009,bjornsonReproducibleResearchBest2019} do not account for all the obstacles to reproducibility of the increasingly complex and large-scale computational experiments. In addition to the complexity of modern computational experiments, these obstacles include the potential for human errors and the rapidly evolving technological landscape that is changing at an unprecedented rate. This article complements the existing works by explicitly focusing on these and related obstacles for computational reproducibility and, in contrast to the discussion in \cite{vandewalleReproducibleResearchSignal2009,bjornsonReproducibleResearchBest2019}, advocates that researchers should plan for the computational reproducibility of their experiments long before any code is written.}

\revise{In summary, although works such as \cite{stoddenImplementingReproducibleResearch2014,vandewalleReproducibleResearchSignal2009,bjornsonReproducibleResearchBest2019} have helped researchers understand the importance of making computational experiments reproducible, the lack of a clear set of standards and tools makes it difficult to incorporate good reproducibility practices in most labs.} It is in this regard that we aim to present signal processing researchers with a set of practical tools and strategies that can help mitigate many of the obstacles to producing reproducible computational experiments.

\subsection{Why \revise{Computational} Reproducibility}
Making \revise{computational} experiments reproducible is a necessary step for ensuring the credibility of the conclusions made from a research study. If researchers are regularly unable to validate the \revise{computational} results in a study, it becomes impossible to investigate whether or not the latest results presented in a research paper are indeed state-of-the-art. \revise{For example, a group of researchers recently published work that presented a new recurrent neural network architecture for language modelling which appeared to achieve state-of-the-art performance in terms of \emph{perplexity} on the Penn Treebank dataset. However, after carefully controlling for hyperparameters, a different group of researchers found that the traditional long short-term memory (LSTM) recurrent neural network model achieved better perplexity on the same dataset, contrary to what was found by the first group. This specific example demonstrates the significance of hyperparameters and other meta data in assessing the performance of new machine learning algorithms. The claims of state-of-the-art performance in particular are directly related to these incidental details and without the ability to reproduce previous computational results it can become difficult to discern the true novelty of new research findings.}

\revise{This example, which is far from a rarity, illustrates the broader point that the field of computational sciences cannot truly advance until we are confident in the past progress.} Not only does ensuring \revise{computational} reproducibility protect the integrity of the research, it also allows fellow researchers to develop their own experiments quickly by utilizing code written by others investigating the same problem. In addition, implementing reproducible \revise{computational} experiments has numerous benefits for the researchers themselves \cite{markowetzFiveSelfishReasons2015}. By utilizing the techniques outlined in this article and making the experiments as transparent and detailed as possible, one can avoid catastrophes such as a hard drive that gets corrupted, resulting in a loss of all data and source code, or a loss of older promising results due to a bug that was introduced in the code later on. Additionally, old research problems often get revisited and improved upon; thus having reproducible code for the experiments in the original study can save hours of frustration for the researcher(s) trying to reproduce old \revise{computational} results, before embarking on the new research. Finally, making one's computational experiments reproducible can add value to the research itself; by lowering the barrier of entry for other researchers to engage and expand on the experiments, it makes the research more accessible to the community, which can in turn lead to more impactful work \cite{piwowarSharingDetailedResearch2007}.

While the importance of \revise{reproducible research in general and computational reproducibility in particular} is obvious to many, the exact tools and techniques that one must utilize when building \revise{computational} experiments to ensure that they are reproducibile for the foreseeable future after publication are still not very clear. In particular, a common misconception is that simply publishing all the code and data used to obtain the results makes \revise{computational} experiments reproducible. But this is almost never the case, as the raw source code and data along with the paper alone cannot give enough details on how to run the experiment, the necessary dependencies, or the required computational power. It is in this regard that we worked on making two experiments from two different research projects here at the INSPIRE Lab reproducible and, in the process, investigated the best ways to create reproducible experiments organically, taking into consideration the extra overhead that comes with such an endeavour. The best techniques and tools were considered in light of the fact that one's goal as a \revise{computational researcher is typically} to conduct and disseminate research and not maintain or develop commercial software. By sharing our experiences and best techniques with the readers of the IEEE Signal Processing Magazine, it is our hope that we can go beyond just highlighting the importance of \revise{computational} reproducibility by providing a clear and practical guide for developing experiments in a reproducible manner.

We have organized the rest of the article as follows. First, the common pitfalls to \revise{computational} reproducibility are explained. Next, the standards for \revise{computational} reproducibility are discussed. The last three sections discuss solutions and tools that can be utilized to avoid the reproducibility pitfalls discussed earlier. These topics include version control for organizing a dynamic and collaborative codebase, package managers for handling multiple dependencies and finally, techniques on how to eventually share the code accompanying the \revise{computational} experiments.

\section{The \revise{Computational} Reproducibility Pitfalls}
\label{sec: pitfalls}
Most papers in the field of signal processing and machine learning tend to include a section at the end where the authors explain their computational experiments along with figures that provide some evidence that the proposed research has practical implications. However, due to the limitations in space and in the interest of conciseness, this section of the research papers cannot provide all the necessary details to reproduce the results. Even when \revise{computational} experiments are only meant to give some justification for a rigorous theory, these results are important and should be explained as clearly as the theorems and proofs on which they are based. While publishing all the source code used for the \revise{computational} experiments is a step in the right direction and may seem to fill in all the missing gaps from the research paper, it is often still not enough to completely reproduce the original results \cite{gundersen2018state}. \revise{This is in part due to the fact that most researchers are not trained on how to write clean, maintainable code in a collaborative setting \cite{alnoamany2018towards}. This lack of training has the potential to result in highly disorganized code that is difficult to parse and understand by an independent researcher.}

Another \revise{potential pitfall we identified, which can seem benign for small experiments}, is that researchers have multiple versions of the same \revise{computational} experiment with slightly different code, which makes it impossible to know which was the one used in obtaining the reported results. Thus, attempting to reproduce the final set of results requires one to first run each version of the codebase individually until they produce the desired results. On large \revise{computational} experiments that take days to finish, this is of course impractical. \revise{Another issue is that researchers typically do not describe in enough detail the dependencies needed for running the code used for the experiment \cite{gundersen2018state}}. Even when the dependencies are mentioned, researchers \revise{might} omit the exact version that was used when originally obtaining results, which could make \revise{computational} reproducibility impossible for those attempting to run the code on a different machine. For example, if the original code used a certain feature from a library that has since been deprecated in later versions, those running the code using the latest version of the library cannot reproduce the results, and may not know that they need to downgrade their dependency. Therefore, there must be a way for the original researchers to preserve and share the exact computational environment used when generating results in order to share it with others looking to run their code to reproduce results. Additionally, as mentioned earlier, the source code alone does not provide instructions on how to run the \revise{computational} experiments nor the order in which the scripts should be executed. Another piece of information that \revise{is rarely mentioned in enough detail when sharing the code is the computational power needed for an experiment \cite{gundersen2018state} and the time it takes to finish executing.} This information is necessary for those trying to reproduce the \revise{computational results} as they must first verify that they are equipped with the right amount of computational power to run the code.

Even when these pitfalls are accounted for, there is still the challenge of sharing the necessary meta data accompanying the experiments. The description of the experiments present in most research papers simply cannot encapsulate all the necessary meta data needed to successfully reproduce \revise{computational} results. For instance, while the authors typically mention the dataset being used in a particular experiment, there may exist some ambiguity about the exact source of the data. And this can be true even for established benchmark datasets such as the ``House'' image or the MNIST dataset\cite{lecunGradientbasedLearningApplied1998}. While the authors might believe that these are ``standard'' datasets, it is often the case that different versions of these datasets are circulating around the internet, each with slight variations that may not be immediately noticeable but can yield different results when used for the same experiment \cite{vandewalleReproducibleResearchSignal2009}. Even if the sources of the datasets are explained, the preprocessing steps performed on the data can be vital to obtaining the published \revise{computational} results, but these \revise{might not be} thoroughly explained in research papers. All of these seemingly benign or superfluous details can have an effect on the \revise{computational} results produced from the experiment that others may be trying to reproduce. Moreover, for \revise{computational} experiments that use synthetic data, the way in which the data was produced \revise{may not be} explained in enough detail. Furthermore, signal processing and machine learning algorithms typically rely on finely tuned hyperparameters that, when changed, can also give different results and even invalidate the conclusions made in the paper, \revise{yet the hyperparameters or the methods by which they are found are not always described in enough detail \cite{gundersen2018state}}. For example, most machine learning experiments make use of cross-validation to find optimal hyperparameters and while the researchers may mention this detail, they might omit information about how exactly the dataset was split up. This in turn could lead to different hyperparameters when the experiment is run by someone trying to reproduce the original results.

Finally, most experiments rely on random number generators somewhere in the codebase. An example of this would be the stochastic gradient descent (SGD) algorithm \cite{robbins1951stochastic}, which is a popular method for large-scale training in machine learning. In each iteration of the vanilla SGD, a random sample from the dataset is used to compute the gradient of the loss function and update the parameters of interest. As a result, the rate of convergence and the values of the optimized parameters depend on the order in which the samples were selected \cite{bottou2012stochastic} and will be different if the randomly chosen sample in each iteration is not consistent every time the experiment is run. By not \revise{saving} the random seed in the code \revise{as part of the experimental meta data}, the sequence of randomly chosen samples will vary every time someone tries to run the experiment, making it almost impossible to \revise{exactly} reproduce the original \revise{computational} results each time the \revise{codebase} is run.

In the rest of this article we discuss what it means for a \revise{computational} experiment to be \revise{computationally} reproducible and the potential techniques for overcoming each of the pitfalls discussed. While the suggested methodologies for creating \revise{computational} experiments may incur some additional work for the researcher, it strikes a good balance between ensuring reproducible results and becoming an obstacle to further research. Furthermore, through practice over time, it is our hope that the tools and techniques discussed below will become commonplace for researchers, giving them the ability to naturally create \revise{computional} experiments that are readily \revise{computationally} reproducible.

\begin{table}[h]
    \caption{A table of the common pitfalls discussed in this section and where to find their respective remedies throughout this article.}
    \label{tab:my_label}
    \centering
    \begin{tabularx}{\textwidth} {
      | >{\centering\arraybackslash}X
      | >{\raggedright\arraybackslash}X
      | >{\centering\arraybackslash}X | }
    \hline
    Pitfall & Short description & Remedy \\
    \hline
    1 & Disorganized codebases and multiple versions. & \hyperref[sec:developing]{Section IV} \\
    \hline
    2 & Differences in computational environments and failure to disseminate necessary dependencies. & \hyperref[sec:dependencies]{Section V} \\
    \hline
    3 & Missing critical meta data such as hyperparameters, computing power and dataset sources. & \hyperref[sec:sharing]{Section VI} \\
    \hline
    \end{tabularx}

\end{table}

\section{The Goals for Reproducibility}
\label{sec: goals}
There has been a lot of discussion across many domains, including within the signal processing and machine learning community, about how to successfully \revise{make computational experiments reproducible}. However, much of the work discussing \revise{computational} reproducibility tends to advocate for new software tools that can be used to easily publish computational experiments that are reproducible. One such example is the ``Whole Tale''\cite{brinckmanComputingEnvironmentsReproducibility2019}, a platform that enables researchers to create \emph{tales}---executable research objects that capture data, code, and computational environments---for the \revise{computational} reproducibility of the research findings. While tools like this may seem promising, they have limitations. The first is that these reproducibility tools attempt to encapsulate the whole process of running an experiment from preprocessing to plotting, treating the \revise{computational} experiment as a blackbox and this in turn leaves the actual code in a state that is difficult to interpret for those who might be interested in digging deeper or expanding upon it. Additionally, due to the fact that they attempt to encapsulate every component of an experiment, they tend to be highly inflexibility and may not be appropriate for every \revise{computational} experiment, making the process of creating \revise{computationally} reproducible experiments even more cumbersome. This is especially true for the research labs that tend to focus more on the theoretical and algorithmic aspects of signal processing and machine learning, rather than the applied aspects, and therefore provide only simple \revise{computational} experiments to give empirical justifications for their claims. In such labs the actual codebases for the computational experiments are not very large and they tend to focus narrowly on a newly proposed idea, insight, or algorithm, which is in contrast to large data analysis projects found in other computational sciences. In addition, the encapsulation-based reproducibility tools are often funded by grant money and are typically only maintained by a single lab; thus when the grant money runs out, there is no guarantee that the tool will be maintained and it may become obsolete within a few short years\cite{konkolPublishingComputationalResearch2020}. Therefore relying on these \revise{computational} reproducibility tools would require one to relearn a new platform for publishing their experiments in a \revise{computationally} reproducible way every time an old one gets abandoned. \revise{In summary, while an encapsulation tool could in theory be the optimal solution to the computational reproducibility crisis, \revise{there currently does not exist any widely adopted off-the-shelf tool} that overcomes the issues highlighted here.}

Because of the aforementioned reasons, the focus of this article is not on finding \revise{or creating} the best \revise{computational} reproducibility software that can automatically encapsulate all the components of the \revise{computational} experiment. Rather, the goal is to find how one can achieve \revise{computational} reproducibility by relying on robust open-source tools, used across both industry and academia. This involves formulating a methodology for developing \revise{computational} experiments such that the reproducible codebase supporting a new research finding can be released in parallel with the publication of the paper. To be more precise, the methodology should enable any independent researcher studying a similar problem to obtain, understand, and easily run the code used in the \revise{computational} experiments in order to reproduce the exact same figures, plots or values without enduring a painstaking trial and error process. Furthermore, this should be possible without the need to ever contact the researchers responsible for the original \revise{computational} experiment.

\revise{The goals of computational reproducibility espoused in this article also emphasize the importance of making experiments computationally reproducible throughout their development, since trying to retroactively make the experiments computationally reproducible after the results have been obtained and published is usually a much more difficult task. If computational reproducibility is only attempted after publication, the researcher is likely to have already fallen into one of the pitfalls mentioned earlier, and must struggle with first reproducing their own results, perhaps unsuccessfully, before being able to share the code with others.} While it may be cumbersome at first, after researchers get accustomed to the tools and techniques presented in this article, the process of making experiments \revise{computationally} reproducible should incur minimal overhead on their research. In particular, it is our hope that the solutions put forth in this article can help overcome all the current hurdles to seamless \revise{computational} reproducibility. These solutions are discussed under three main themes in the following. Version control systems keep track of changes in the codebase as well as eliminate the issue of multiple concurrent versions of the experiments. Therefore, in order to support organized and understandable codebases, we first present the best tools for implementing version control. Next, in an effort to ensure that dependencies are accurately and easily shared, we discuss the simplest tools for dependency management that allow the researchers to disseminate their exact computational environment to others. Finally, in order to make the codebase easily explorable and provide thorough instructions on how to \revise{computationally} reproduce the experiments, we offer some advice on the best ways to document and publish the code for sharing with the rest of the research community.

\section{Managing a Developing Experiment}
\label{sec:developing}
Research is all about taking incremental steps towards a result. This is true when trying to prove theorems as well as in developing computational experiments. In practice the algorithms that get applied to datasets always need some level of tuning in order to run the \revise{computational} experiments correctly or obtain the best possible results. Oftentimes one wants to investigate how changing a certain piece of the code (e.g., a specific hyperparameter) alters the results without losing the current version of the code they have. A na\"ive solution to this problem is to make a copy of the source code with the desired changes without deleting the original. For example, a file named \verb|my_algo.py| containing an implementation of the main algorithm for the experiment might get copied and renamed to \verb|my_algo_1.py| with a few subtle changes made within it. Later the researchers may be interested in changing things a little differently but still keep \verb|my_algo.py| and \verb|my_algo_1.py| just in case the new version performs worse, so they create another file named \verb|my_algo_2.py|, and this pattern repeats for multiple versions of the file. Finally, a few months after the work is published, someone might ask about how the figures were produced and the original researchers are left scrambling through up to a dozen different versions of the codebase, trying to find the one that actually produced the right results. Clearly this is not a good solution and can quickly become an unwieldy situation, forcing one to re-run multiple versions of the codebase while trying to find the correct one. Though this ad-hoc approach may have been feasible for small \revise{computational} experiments that only take seconds or minutes to finish computing, with the rise in larger datasets and higher wall-clock time per \revise{computational} experiment, re-running multiple versions of the experiments until one finds the right version can take days or even weeks. This is the precise problem we found in one of our codebases that we tried to \revise{computationally} reproduce, which had multiple versions of the same experiment. Even the original researchers on the project were not able to recall the version  that corresponded to figures in the published work. Given that some of these \revise{computational} experiments took up to five days to finish running, finding the version that reproduced the original plots took us weeks.

Another common hurdle that researchers must overcome is trying to develop \revise{computational} experiments with collaborators. Due to the nature of computational research, developing \revise{computational} experiments collaboratively can be done without having to physically share the same computational resources, lab or even the same country. While this certainly makes collaboration easier compared to other scientific domains, collaborating in the development of a codebase is still not a simple task. Some of the current solutions that researchers employ are faulty and have many drawbacks. \revise{For instance a researcher may treat source code like any other file and rely on tools such as Google Drive\cite{CloudStorageWork}, DropBox\cite{HomeDropbox}, Microsoft OneDrive\cite{PersonalCloudStorage} and email to share their code amongst collaborators.} While these tools are great for sharing images, documents and other forms of media, they are not the best tools for sharing a developing codebase, as they require the collaborators to constantly download and then upload the codebase every time a change has been made. There is also a responsibility on the one making the changes to alert all other collaborators in order for them to re-sync their codebases. In short, this mode of collaboration can actually slow down collaboration and is prone to errors because it makes it very cumbersome to track changes made in the codebase and ensure that everyone is using the latest version of the codebase.

One of the best solutions to both of the aforementioned problems is an effective use of a good version control system, the most popular being \verb|git|, a free and open-source distributed version control system known for its efficiency compared to other version control systems, making it easy to keep track of all the changes being made to an experiment throughout the development process. Due to its popularity and utility \verb|git| integrates well with many online code hosting services, such as GitHub\cite{BuildSoftwareBettera}, BitBucket\cite{atlassianBitbucketGitSolution}, and GitLab\cite{DevOpsPlatformDelivered}. The advantage of \verb|git| is that it allows one to track every change that has been made to any file in the codebase, often referred to as the \emph{repository}. Once one of the files in the repository being tracked has been changed, the researcher can then move that file or set of files to the \emph{staging} area. From the staging area the changes are then \emph{committed}, which assigns the current state of the repository a unique SHA-1 hash and saves it in a history database so that it can be compared to future versions of the codebase or recovered when things go wrong. These commits also contain information of who made the changes and are accompanied by a short comment or message to explain what change were made and why. By committing regularly, a researcher is able to traverse through all changes of the codebase and see precisely which lines of code were changed, by whom they were changed, and a short description of why they were changed. The \verb|git| branching feature also lends itself very well to experimenting and trying out different parameters or techniques. By \emph{branching}, one can create an alternate tracking history of the repository starting from the current commit, and from there one can edit the code and run the experiments without affecting the current implementation on the main branch.

However, \verb|git| truly shines in a collaborative setting. This is primarily due to its design as a distributed version control system, which means that it provides each collaborator a full commit history of the entire repository. When used jointly with a code hosting platform, such as GitHub\cite{BuildSoftwareBettera}, BitBucket\cite{atlassianBitbucketGitSolution}, or GitLab\cite{DevOpsPlatformDelivered}, it allows researchers to upload their repository to the internet and \emph{pull} down changes from the hosting service to their local environment every time they intend to develop on the codebase. This ability to pull down the changes is substantially better than the traditional approach of re-downloading a codebase from cloud storage providers because it eliminates the need to ask collaborators about what was altered or alert other collaborators of the changes implemented. This is because every commit that was made is already documented by the commit message accompanying it, along with the name of the person who made the commit. Furthermore, due to the efficient design of \verb|git|, pulling down changes from a repository only takes seconds even if numerous changes were made; this is in contrast to re-downloading an entire codebase from a cloud storage provider, which can take minutes every time. Additionally git is intelligently designed to handle merging the changes between the current version of the code on one's local machine and the updated codebase online, making the process very seamless and facilitating efficient asynchronous collaboration.

While some researchers may argue that \verb|git| is a complex tool which ultimately impedes their productivity, this is only temporary and we believe that that it is worth the investment for computational researchers. Furthermore, most researchers in machine learning and signal processing collaborate extensively with industry where version control is expected to be used in maintaining any codebase. Perhaps the most appealing feature of using \verb|git| for version control over other alternatives is its widespread use among programmers. Over the years, this has resulted in numerous resources all over the internet in the form of videos, blog posts \cite{UnderstandingGitConceptually}, and books\cite{chaconProGit2014}. Additionally, online communities such as StackOverflow \cite{StackOverflowWhere} provide answers to nearly any confusion one may have about \verb|git| and its features. There are also a number of graphical interface tools for using \verb|git| for those not yet comfortable using the command line. While \verb|git| is an extremely flexible tool and can be a tremendous aid in writing organized and manageable code for experiments across multiple researchers working together, it is only effective when  utilized properly. Researchers must ensure that they are committing changes frequently and are providing providing concise and accurate descriptions of the changes made for easy reference as the codebase is developed.

\begin{mdframed}[style=MyFrame]
\begin{center}
    \includegraphics[width=13cm]{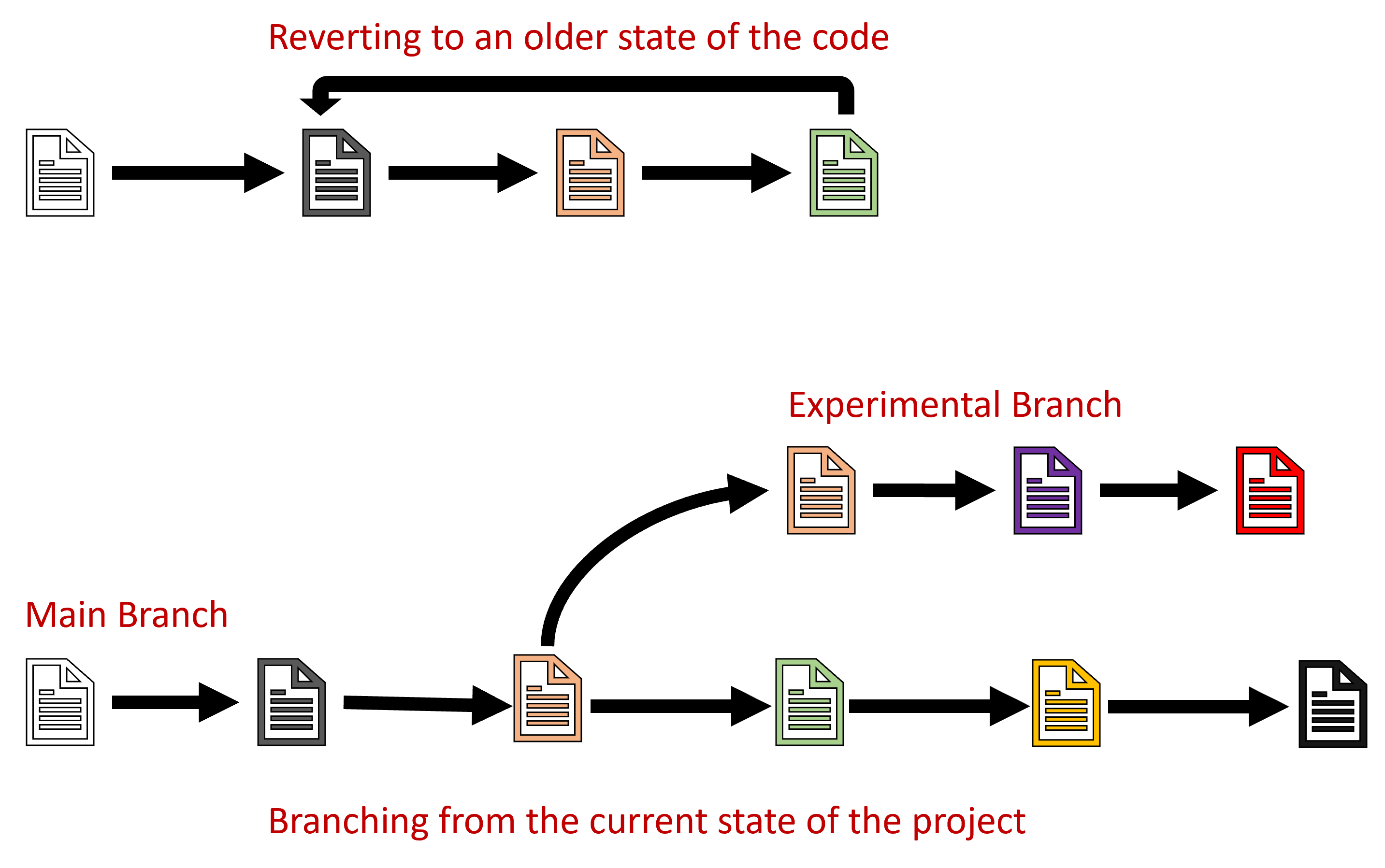}
\end{center}
\noindent\emph{Version Control}: By effectively using version control and making frequent commits, researchers do not have to save multiple versions of the same file in their codebase, because they are always able to easily revert back to an earlier commit. With version control, specifically \verb|git|, one is also free to pursue multiple versions of the experiment without impinging on each other. This is most effectively done by branching, which allows one to create multiple version histories.
\end{mdframed}

Going beyond version control, the next important step for ensuring reproducibility is to make sure that those trying to reproduce the code are able to capture the same exact computational environment as the one used by the original researchers. This is the topic of our next discussion.

\section{Managing Dependencies}
\label{sec:dependencies}
\revise{One of the simplest ways to ensure a robust and reproducible codebase is to try and minimize the use of external packages and libraries in the code. However, due to the immense utility of modern software libraries, the codebases for modern signal processing and machine learning experiments are seldom developed without the use of several external libraries. These include popular scientific programming and machine learning libraries such as NumPy\cite{harrisArrayProgrammingNumPy2020}, Scikit-learn\cite{pedregosaScikitlearnMachineLearning2011}, TensorFlow \cite{abadiTensorFlowLargeScaleMachine}, CVX\cite{cvx}, and Tensor Toolbox \cite{TTMATLAB}}. Although there are multiple programming languages available for scientific computing, such as R\cite{rcoreteamLanguageEnvironmentStatistical2020}, Julia\cite{bezansonJuliaFreshApproach2017}, and MATLAB\cite{mathworksMATLAB}, we focus the discussion specifically on the Python programming language as it is free, open source and widely popular in the machine learning community. However, the techniques discussed in this section can still be utilized for other popular programming languages.

Without first knowing which exact dependencies need to be installed on a researcher’s machine, it becomes impossible to reproduce results. The reproducer must not only be aware of what dependencies are being used, but also the precise version being used at the time when the results were originally generated. One popular way to encapsulate the original researcher's computational environment perfectly is by using Docker\cite{merkelDockerLightweightLinux2014}, as suggested in \cite{boettigerIntroductionDockerReproducible}. While Docker is a powerful containerization tool and it does indeed solve the problem of dependency management by encapsulating the original researcher's computational environment into isolated \emph{containers}, it could be too much of an overhead for researchers who have no prior experience using Docker or managing large software projects. This is especially true in labs and research groups where researchers focus primarily on the mathematical and algorithmic aspects of their research and only utilize a small codebase for experiments, which relies on only a few dependencies. In these \revise{computational} experiments one typically does not attempt to synthesize and manage multiple programming languages, frameworks and dependencies, as is often the case in real-world commercial applications, making Docker-based solutions an overkill.

Therefore, a more appropriate tool would be a simple and light dependency manager similar to \verb|pip|, the default dependency manager for Python. But the most fitting environment and dependency manager for computational experiments is \verb|conda|\cite{CondaCondaPost8}, an open-source environment and package management tool. Although \verb|conda| can be used independently in any codebase, it is automatically included in the Anaconda \cite{anaconda} distribution of Python, which is already the distribution of choice for many computational researchers. Unlike other package managers, \verb|conda| is specifically designed to easily manage the dependencies most commonly encountered in scientific computing, and overcomes the many shortcomings of \verb|pip|. One major advantage of using \verb|conda| is that when one attempts to installs a new package, \verb|conda| ensures that all the requirements for this new package are met before adding it, and if this is not the case an error is shown immediately with steps on how to rectify the issue. This is contrary to \verb|pip|, which does not check this condition before installing a new package and can result in unexpected errors later during development. However, \verb|conda|'s greatest advantage is allowing one to create independent \emph{virtual environments} for each experiment, so that all projects do not share the same global dependencies. This in turn ensures that each environment contains only the packages that are absolutely needed for the current experiment associated with it and nothing more. Creating these virtual environments is also important because when the version of a package from one project gets upgraded on the researcher's machine, it will not interfere with the current version of that same package in the other environments, allowing the original computational environment on which the experiment was carried out to be preserved. Finally, while Python's standard library does include its own virtual environment manager through the \verb|venv| module, \verb|conda| is unique in that it allows one to create environments with different versions of Python itself.

\begin{mdframed}[style=MyFrame]
\begin{center}
    \includegraphics[width=13cm]{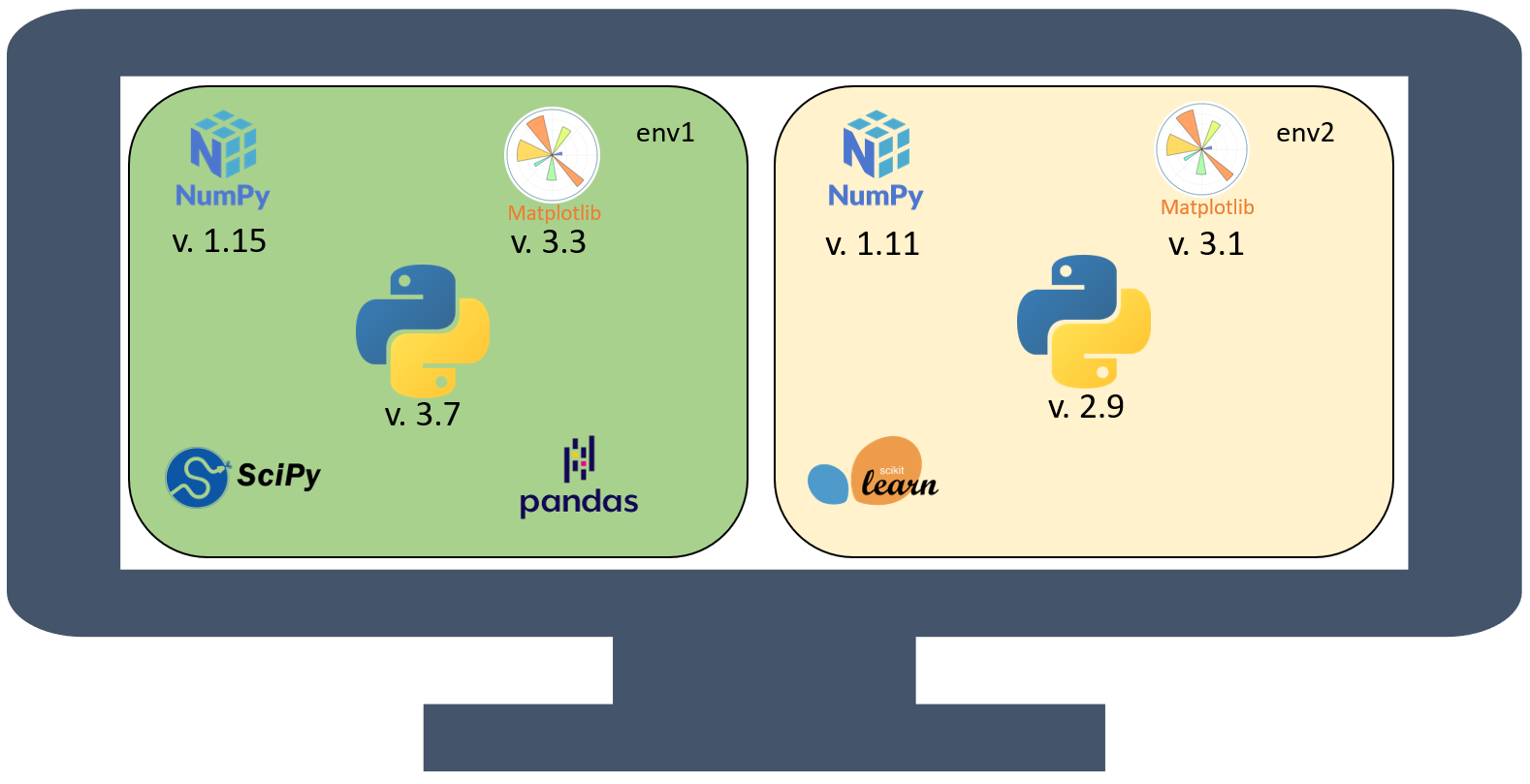}
\end{center}

\noindent\emph{Conda Environments}: A conceptual diagram that illustrates the usefulness of conda environments for reproducible research. On the left, ``env1'' is a python environment with four different libraries commonly used in machine learning experiments. On the right, the environment ``env2'' has some of the same libraries but with different version numbers and a different version of python itself; this could, for example, be an environment for an older variant of the experiment. Both these environments are independent of each other and preserve the computational environment that was originally used to produce experimental results. They can also be easily exported and shared with outside researchers through an \verb|environment.yml| file. \revise{The environment can be reconstructed on a new computer using the command} \verb|``conda env -f environment.yml''|. %
\end{mdframed}

\begin{figure}[t]
    \centering
    \includegraphics[width=\textwidth]{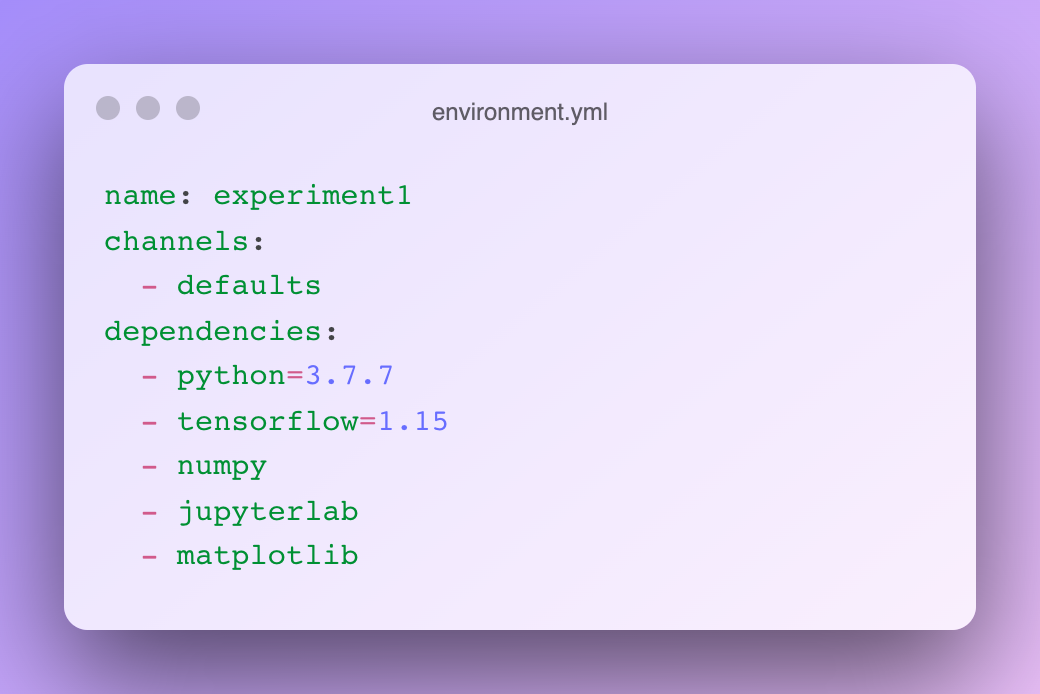}
    \cprotect\caption{\linespread{1}\selectfont\footnotesize An example of an \verb|environment.yml| file with common libraries and dependencies used in typical machine learning experiments.}
    \label{fig:environment.yml}
\end{figure}

Based on the above, we recommend using \verb|conda| for managing dependencies in reproducible experiments. This involves creating an \verb|environment.yml| (or a similarly named) file for each project, which specifies the necessary dependencies, and then constructing a new environment \revise{based on the specifications in this file, which can be done with a simple command}. This should be done before any code for the experiment is written in order to ensure no dependencies are unaccounted for. If one wants to later add a dependency they can do so by adding it to their \verb|environment.yml| file and updating their environment according to the new additions made to the file. Then, when another independent researcher wishes to run the experiment, they can quickly and easily reconstruct the same \verb|conda| environment with all the necessary dependencies by simply using the \verb|environment.yml| file. \revise{Additionally, one can specify the exact version number for the important dependencies in the ``environment.yml`` file to ensure that when the ``conda`` environment gets reconstructed on another machine (potentially with another operating system) the same dependency version is used.} Also, for those who prefer some other dependency manager as opposed to \verb|conda|, they can still inspect the plain text yaml file to see all the necessary dependencies with their respective versions. \revise{A typical ``environment.yml`` for a machine learning experiment is shown in  Figure~\ref{fig:environment.yml}, demonstrating its readability and effectiveness in organizing dependencies. Here we have \emph{pinned} the version for TensorFlow to ensure that this same version is always preserved when the environment is reconstructed on a different machine.} By utilizing this tool, managing dependencies becomes very straightforward and makes it possible for others to replicate the same computational environment that one had used when originally running the experiment, without the need to resort to trial and error or to contact the original researchers.

Proper version control and dependency management tools are critical first steps to promoting reproducibility throughout development of an experiment. However, ensuring reproducibility for the foreseeable future is made most probable through properly shared code and data, detailed documentation and thoughtful organization of the codebase developed. We now shift our focus to this aspect of reproducibility.

\section{Sharing Code}
\label{sec:sharing}
While version control and dependency management make it possible for others to run the code and help to keep the codebase organized, this does not provide any information about the order in which the scripts should be executed and what computational resources were utilized in obtaining the original results. Furthermore, when researchers look to reproduce other's experiments, they do not necessarily want to reproduce each and every figure in the paper. This is especially true for papers with multiple experiments. Sometimes they are only interested in a specific plot or the implementation of a particular algorithm that was described in the original work. Therefore it is important to ensure that one's experiments do not get crammed into a single source file that includes preprocessing, algorithmic implementations, analysis and visualization. Indeed, by making the codebase modular and organizing the project structure carefully, as discussed in the following, it becomes significantly easier for others to download and inspect what they want to reproduce from the codebase without too much digging.

\textbf{Project Structure:} Typically, a computational experiment in signal processing and machine learning can be broken down into three parts:
\begin{enumerate}
    \item Preprocessing of data or generation of synthetic data.
    \item Execution of the actual experiment on the data.
    \item Analysis of results and generation of figures.
\end{enumerate}
It is often advantageous to keep the project folder organized in a similar fashion by creating separate scripts, or even subfolders, for each of these tasks. Keeping the codebase modular and structured in this manner allows others looking to reproduce a particular set of results to do so without needing to analyze or execute the entire codebase. For example, in our lab's experiments on learning mixtures of separable dictionaries for tensor data \cite{joseph_shenouda_2020_3901877}, we found it necessary to split the experiments up into different subfolders as the original work \cite{learningghassemi} compared four different algorithms on both synthetic and real datasets.  Object-oriented design within the code can also be helpful as this increases the modularity of the experiments and allows extensions to be developed by other researchers naturally. Last, but not the least, the data used in the experiments must also be shared correctly; this involves ensuring that not only the raw data get shared but also all the intermediate steps or scripts used to preprocess the data before using them in the experiments are shared.

\revise{\textbf{Coding Standards:}} Additionally, it is worth following a set of guidelines for clear and concise comments that can describe every function or class definition in the codebase. For Python the convention is to add a \emph{docstring} \cite{PEP257Docstring} to each module, as many text editors and integrated development environments (IDEs) often search for these in the project files to allow programmers to quickly inspect what a function is doing without actually going to its destination in the source code. \revise{This should be accompanied by standard coding practices, such as proper variable naming conventions and indentations, which can make the codebase easier to read and interpret by others.}

\textbf{Documentation:} The most important thing that must accompany the project being shared is a detailed README file. We recommend that the README contain all the following elements:
\begin{enumerate}
    \item Brief description of the project and related paper(s).
    \item Steps to install necessary dependencies, e.g., via \verb|conda| and \verb|environment.yml| file.
    \item Computational resources used to run the experiments along with their respective runtimes.
    \item Scripts used to preprocess the data, in the case of real data, or generate them for the case of synthetic data.
    \item Description of which scripts one should run for which experiments and how to then plot the results.
\end{enumerate}
We found that each of these pieces of information were necessary when attempting to reproduce the results from a codebase. Such documentation that accompanies the code provides a full picture of how an experiment should be run and how one actually acquires the desired results without having to spend much time laboring through every detail in the code.

\textbf{Publicization:} The way in which the code is publicized also needs careful consideration to ensure ease of access and, most importantly, permanency. Typically experiments are shared by hosting the accompanying source code on the lab’s website or on a code hosting platform, such as GitHub, GitLab or BitBucket. However, posting it on the lab’s website is unlikely to be the most robust solution; for instance, if a researcher leaves his or her position, their website often disappears along with the code. Code hosting services, however, are linked to an account and repositories can be easily transferred to different owners with minimal hassle. Sharing the code this way also allows others to fork the repository and develop further on it. Additionally, others can add their comments or questions about the code in the form of ``Issues.'' These discussions become public for anyone else viewing the repository as well, which can eliminate redundant questions that the community may have. One of the biggest issues with both of these solutions is that they are still dependent on the organization hosting the website. This means that if the hosting website were to ever disappear, the code hosted on it would go with it. But permanency on the internet is necessary for ensuring that the results are reproducible for the forseeable future. One way to ensure permanency is to assign the codebase a digital object identifier (DOI) to give it a permanent presence on the internet, which can be done with tools such as Zenodo\cite{ZenodoResearchShared}.

\section{Conclusion}
\label{sec:conclusion}
While there has been much discussion in the literature about the reproducibility crisis in computational research, not enough emphasis has been given on the best practices and techniques to solve this problem with established tools. The solutions to \revise{computational} reproducibility have been especially unaddressed for the more theoretically bent research groups within the signal processing and machine learning community that typically develop smaller computational experiments compared to other computational sciences. In this article, we presented the main pitfalls to achieving reproducible experiments and then provided common tools and techniques that can be used to overcome each of those pitfalls, while bearing in mind that making experiments reproducible can entail extra effort that may divert our attention away from our primary task of research. By utilizing the right tools for version control and dependency management as well as careful structuring and documentation when sharing the codebase, we can work towards ensuring that every \revise{computational} experiment in our research is readily reproducible.

\section*{Acknowledgements}
The authors gratefully acknowledge the support of the NSF (CCF-1453073, CCF-1907658, CCF-1910110, OAC-1940074) and the ARO (W911NF-17-1-0546, W911NF-21-1-0301).

\section*{Short Biographies}
\noindent\textbf{\emph{Joseph Shenouda}} (jshenouda@wisc.edu) received his B.S. degree in Electrical \& Computer Engineering in 2021 from Rutgers University--New Brunswick. He is currently a graduate student at the University of Wisconsin--Madison in the Electrical \& Computer Engineering department. His research interests include graph signal processing and machine learning.

\noindent\textbf{\emph{Waheed U. Bajwa}} (waheed.bajwa@rutgers.edu) has been with Rutgers University--New Brunswick, NJ since 2011, where he is currently an associate professor in the Departments of Electrical \& Computer Engineering and Statistics. His research interests include statistical signal processing, high-dimensional statistics, and machine learning. He is a senior member of the IEEE.


\end{document}